\documentclass[12pt,preprint]{aastex}

\slugcomment{Accepted for publication in \textbf{Ap\&SS}}

\begin{document}


\title{Investigating thermal evolution of the self-gravitating one dimensional molecular cloud by smoothed particle hydrodynamics}


\author{Mohsen Nejad-Asghar\altaffilmark{1,2} and Diego Molteni\altaffilmark{3}}

\affil{$^1$Department of Physics, Damghan University of Basic
Sciences, Damghan, Iran}

\affil{$^2$Research Institute for Astronomy and Astrophysics of
Maragha, Maragha, Iran}

\affil{$^3$Dipartimento di Fisica e Tecnologie Relative,
Universita di Palermo, Viale delle Scienze, 90128, Palermo, Italy}



\begin{abstract}
The heating of the ion-neutral (or ambipolar) diffusion may
affect the thermal phases of the molecular clouds. We present an
investigation on the effect of this heating mechanism in the
thermal instability of the molecular clouds. A weakly ionized one
dimensional slab geometry, which is allowed for self-gravity and
ambipolar diffusion, is chosen to study its thermal phases. We
use the thermodynamic evolution of the slab to obtain the regions
where slab cloud becomes thermally unstable. We investigate this
evolution using the model of ambipolar diffusion with two-fluid
smoothed particle hydrodynamics, as outlined by Hosking \&
Whitworth. Firstly, some parts of the technique are improved to
test the pioneer works on behavior of the ambipolar diffusion in
an isothermal self-gravitating slab. Afterwards, the improved
two-fluid technique is used for thermal evolution of the slab. The
results show that the thermal instability may persist
inhomogeneities with a large density contrast at the intermediate
parts of the cloud. We suggest that this feature may be
responsible for the planet formation in the intermediate regions
of a collapsing molecular cloud and/or may also be relevant to the
formation of star forming dense cores in the clumps.
\end{abstract}





\section{Introduction}

Observations of molecular clouds show the existence of small
density fluctuations threaded by magnetic fields. The recent
progress of the molecular cloud observations has established that
the small and tiny scale structures are very ubiquitous. For
example, Langer et al. (1995) observed small condensations with
size from $0.007$ to $0.021\mathrm{pc}$ in Taurus Molecular
Cloud~1. The mass of these fragments is estimated to be $<
0.01-0.15 M_\odot$. Studies of the time variability of absorption
lines indicates the presence of fluctuations on scales of
$10^{-4} \mathrm{pc}$ (5-50 AU) and masses of $10^{-9}
\mathrm{M_\odot}$ (Boiss\'{e} et al. 2005). At larger scales
(about $10,000$~AU), Pan et al. (2001) find significant
differences in $\mathrm{CN}$, $\mathrm{CH}$, and $\mathrm{CH^+}$
absorption lines. The \textsc{spitzer} experiment has begun
producing higher spatial resolution mid-infrared maps (Churchwell
et al. 2004) and revealed the fine structure of the star forming
regions. The advent of major new facilities in the coming years
should yield several breakthroughs in this field (Andr\'{e} et
al.~2008). A higher resolution observation in future is assuredly
expected to reveal hidden, small structures, as well as
sub-stellar objects with very small mass (Atacama Large
Millimeter Array, \textsc{alma}, becoming partly available in
2011, fully operational at 2013).

The classical problems of cloud fragmentation are relevant to star
formation because they show that there is a preferred scale (Jeans
length) of gravitational instability as soon as one considers an
initial structured region with some length-scale (see, e.g.,
Larson~1985). By the way, it is not easy for the small and tiny
scale structure to form as a result of gravitational instability,
since their size are much smaller than the Jeans length. According
to Langer et al. (1995) indeed, the small scale structures appear
to be gravitationally unbound. This suggests that some
fragmentation mechanisms other than pure Jeans gravitational
instability may be important in the clouds. Molecular clouds are
thought to be supersonically turbulent, since they exhibit
supersonic linewidths (Zuckerman \& Palmer~1974). Base on the
idea of turbulence, some mechanisms to generate substructures
inside molecular clouds have been proposed (Elmegreen 2007). As
well, a further process based on magnetohydrodynamic waves have
been suggested to produce inhomogeneities in the clouds (Folini et
al.~2004). Another basic physical process that could trigger
formation of density fluctuations is thermal instability, because
it acts on time-scales that can be much shorter than the duration
of the turbulent motions (Gilden 1984). The effect of thermal
instability in fragmentation of the clouds is clearly a
competitive process relative to other mechanisms, although all
could operate.

In molecular clouds, a neutral molecular gas is intermixed with an
ionized component that is tied directly to the magnetic field.
Birk~(2000) has used the two-fluid technique to find the thermal
condensation modes in weakly ionized hydrogen plasma. Nejad-Asghar
\& Ghanbari~(2003) studied the effect of linear thermal
instability in a weakly ionized magnetic molecular cloud within
the one-fluid description. Nejad-Asghar (2007) has recently made
the assumption that the molecular cloud is initially an uniform
ensemble which then fragments due to thermal instability. He find
that ambipolar drift heating is inversely proportional to density
and its value, in outer parts of the cloud, can be significantly
larger than the average heating rates of cosmic rays and
turbulent motions. His results show that the isobaric thermal
instability can occur in intermediate regions of the cloud;
therefore it may produce the cloud fragmentation and formation of
the condensations.

The study under the one-fluid approach did not precisely consider
the effect of the ion-neutral friction, since they did not treat
the flow as two fluids which are composed of ions and neutrals.
Then, Fukue \& Kamaya~(2007) revisited the effect of the
ion-neutral friction of the two-fluid on the growth of the linear
thermal instability. Their results indicate that the friction
with the magnetic field affects the morphology and evolution of
the interstellar matter. Falle et al.~(2006) described numerical
calculations using an \textsc{amr} magnetohydrodynamic code that
show that thermal instability may have an important role to play
in the formation of the hierarchical structure of molecular
clouds. Here, we use the two-fluid numerical simulation to study
the evolution of weakly ionized molecular clouds in the nonlinear
thermal instability regime.

Many authors have developed computer codes that attempt to model
ion-neutral diffusion. Black \& Scott~(1982) used a
two-dimensional, deformable-grid algorithm to follow the collapse
of isothermal, non-rotating magnetized cloud. The
three-dimensional work of MacLow et al. (1995) treats the
two-fluid model in a version of the \textsc{zeus}
magnetohydrodynamic code. An algorithm capable of using the
smoothed particle hydrodynamics (SPH) to implement the ambipolar
diffusion in a fully three-dimensional, self-gravitating system
was developed by Hosking \& Whitworth (2004, hereafter HW). They
described the SPH implementation of two-fluid technique that was
tested by modeling the evolution of a dense core, which is
initially thermally supercritical but magnetically subcritical.

In this paper, we firstly re-formulate and improve the two-fluid
SPH implementation of the ambipolar diffusion in an isothermal
self-gravitating one dimensional slab. Afterwards, we use this
two-fluid SPH technique to investigate the nonlinear thermal
evolution of the slab. In this way, the continuum equations of the
self-gravitating slab and its isothermal case are given is
section~2. The two-fluid SPH technique is presented in section~3.
Section~4 devoted to the chosen physical scales, initial
conditions, and the computer experiments to explore the
temperature profile and density fluctuations in the contracting
one dimensional molecular cloud. Finally, section~5 is allocated
to the summary and conclusions with some prospects.

\section{Gas dynamics}
In the absence of reliable information about the structure of the
molecular clouds, it is reasonable to examine simple geometries
such as sphere, disks, cylinder, etc., in order to gain some
insight into the evolution process. Here, we consider an
one-dimensional lightly ionized molecular gas with purely
transverse magnetic field, thus, all variables are functions of
distance $z$ to the center and time $t$ only.

\subsection{The continuum equations}
The exact fraction of the total fluid that is ionized depends upon
many factors (e.g. the neutral density, the cosmic ray ionization
rate, how efficiently ionized metals are depleted on to dust
grains). Here, we use the expression employed by Fiedler \&
Mouschovias (1992), which states that for $10^{8} < n < 10^{15}
\mathrm{m}^{-3}$,
\begin{equation}\label{ionden}
  \rho_i=\epsilon (\rho^{1/2} + \epsilon' \rho^{-2}),
\end{equation}
where in standard ionized equilibrium state, $\epsilon\sim 7.5\times
10^{-15} \mathrm{kg^{1/2}.m^{-3/2}}$ and $\epsilon' \sim 4\times
10^{-44} \mathrm{kg^{5/2}.m^{-15/2}}$ are valid. In reality, the gas
in this case is very weakly ionized, thus, we adopt the
approximation $\rho=\rho_n+\rho_i\approx\rho_n$ in fluid equations.
In this case, the molecular cloud is considered as global neutral
which consists of a mixture of atomic and molecular hydrogen (with
mass fraction $X$), helium (with mass fraction $Y$), and traces of
$\mathrm{CO}$ and other rare molecules, thus, the mean molecular
weight is given by $1/\mu=X/2+Y/4$.

We write the continuity equation of neutral part as its common form
\begin{equation}\label{masscon}
\frac{d \rho}{d t} =- \rho \frac{\partial v}{\partial z},
\end{equation}
while, the relation (\ref{ionden}) is used to determine the ion
density wherever it is required in the two-fluid equations. The
momentum equation of neutral part then becomes
\begin{equation}\label{momcon}
  \frac{d v}{d t} = g - \frac{1}{\rho} \frac{\partial}{\partial z} (p +
  \frac{B^2}{2\mu_0})
\end{equation}
where the gravitational acceleration $g$ obeys the poisson's
equation
\begin{equation}\label{poisson}
  \frac{\partial g}{\partial z} = -4 \pi G \rho,
\end{equation}
and the pressure is given by the ideal gas equation of state
\begin{equation}\label{pressu}
  p=\frac{R}{\mu}\rho T
\end{equation}
where $R$ is the molar gas constant and the temperature $T$ is
approximated the same as for both neutral and ion fluids
($T_i=T_n=T$).

The calculations are usually further simplified by the assumption
of an isothermal equation of state. This last assumption is,
however, unnecessarily crude since molecular gas is expected to
cool and heat rapidly and appropriate cooling and heating
function may be estimated (e.g., Goldsmith~2001). The thermal
energy per unit mass is generally given by
\begin{equation}\label{ut}
u(T)=(\frac{5}{4}X + \frac{3}{8}Y) \frac{k_B T}{ m_H}
\end{equation}
where the mean internal (rotation and vibration) energy of an
$\mathrm{H_2}$ molecule is included. At temperature below $\sim
200\mathrm{K}$, the rotational and vibration degrees of freedom
of molecular hydrogen are not excited, so it actually behaves
like a monatomic gas. The energy equation follows from the first
law of thermodynamics, that is
\begin{equation}\label{energycon}
  \frac{du}{dt} = - \frac{p}{\rho} \frac{\partial v}{\partial z} - \Omega_{(\rho,T)},
\end{equation}
where $\Omega_{(\rho,T)}$ is the net cooling function
\begin{equation}\label{netcool}
  \Omega_{(\rho,T)} \equiv \Lambda_{(n)} (\frac{T}{10 \mathrm{K}})^{\beta_{(n)}} -
  (\Gamma_{CR} + \Gamma_{AD}),
\end{equation}
where $\Gamma_{CR}$ and $\Gamma_{AD}$ are the heating rates due to
cosmic rays and ambipolar diffusion, respectively, and
$\Lambda_{(n)}$ and $\beta_{(n)}$  are the parameters for the gas
cooling function that here we use the polynomial fitting
functions, outlined by Nejad-Asghar~(2007), as follows
\begin{eqnarray}\label{lambda0}
  \nonumber \log\left(\frac{\Lambda_{(n)}}{\mathrm{J.kg^{-1}.s^{-1}}}\right) = -8.98 -
  0.87 (\log \frac{n}{n_0}) \\ - 0.14 (\log \frac{n}{n_0})^2,
\end{eqnarray}
\begin{equation}\label{beta}
  \beta_{(n)} = 3.07 - 0.11 (\log \frac{n}{n_0}) - 0.13 (\log \frac{n}{n_0})^2,
\end{equation}
where $n_0=10^{12} \mathrm{m^{-3}}$. The heating of cosmic rays
is given by (e.g., Goldsmith~2001),
\begin{equation}\label{heatCR0}
  \Gamma_{CR} \approx 3.12 \times 10^{-8}\quad \mathrm{J.kg^{-1}.s^{-1}},
\end{equation}
while the heating due to ambipolar diffusion can be physical
derived by considering the drag force (per unit volume) as follows
\begin{equation}\label{heatAD0}
  \Gamma_{AD} = \frac{\textbf{f}_d.\textbf{v}_d}{\rho},
\end{equation}
where
\begin{equation}\label{dforce}
  \textbf{f}_d = \gamma_{AD} \rho_i \rho \textbf{v}_d,
\end{equation}
where $\gamma_{AD} \sim 3.5 \times 10^{10}
\mathrm{m^3.kg^{-1}.s^{-1}}$ represents the collision drag
coefficient.

The magnetic fields are directly evolved by charged fluid component,
as follows:
\begin{equation}\label{magcon}
  \frac{d B}{d t} = - B \frac{\partial v}{\partial z} +
  \frac{\partial}{\partial z} (B v_d),
\end{equation}
where the last term outlines the ambipolar diffusion effect with
drift velocity,
\begin{equation}\label{drift}
  v_d = -\frac{1}{\gamma_{AD} \epsilon \rho \rho_i} \frac{\partial}{\partial
  z} (\frac{B^2}{2\mu_0}),
\end{equation}
which is obtained by assumption that the pressure and
gravitational force on the charged fluid component are negligible
compared to the Lorentz force because of the low ionization
fraction.

\subsection{Isothermal molecular layer}

In isothermal case, the thermal energy per unit mass (\ref{ut})
is overplus, and the pressure in momentum equation (\ref{momcon})
then becomes $p=a^2\rho$ where $a$ is the isothermal sound speed.
Following the many previous treatments, a further simplification
is possible if we introduce the surface density between mid-plane
and $z>0$ as
\begin{equation}\label{sig}
\sigma \equiv \int_{0}^{z} \rho(z',t)dz'.
\end{equation}
By transformation from $(z,t)$ to $(\sigma, t)$, the drift
velocity is given by
\begin{equation}\label{drift2}
  v_d = -\frac{1}{\gamma_{AD} \epsilon \rho^{1/2}(1+\epsilon' \rho^{-5/2})}\frac{\partial}{\partial
  \sigma} (\frac{B^2}{2\mu_0}),
\end{equation}
and the equation (\ref{magcon}) becomes
\begin{equation}\label{magcon2}
  \frac{\partial}{\partial t} (\frac{B}{\rho})= \frac{1}{\gamma_{AD}\epsilon\mu_0}
  \frac{\partial}{\partial \sigma} (\frac{B^2}{\rho^{1/2}(1+\epsilon' \rho^{-5/2})}\frac{\partial B}{\partial
  \sigma}).
\end{equation}
With the above, field equation (\ref{poisson}) can be integrated
to give
\begin{equation}\label{poisson2}
  g = - 4 \pi G  \sigma,
\end{equation}
while the equation of continuity (\ref{masscon}) and the equation
of motion (\ref{momcon}) take the form
\begin{equation}\label{mascon2}
  \frac{\partial z}{\partial \sigma} = \frac{1}{\rho}
\end{equation}
and
\begin{equation}\label{momcon2}
  \frac{\partial^2 z}{\partial t^2} = -4 \pi G \sigma
   - \frac{\partial}{\partial \sigma} (a^2 \rho+ \frac{B^2}{2\mu_0}),
\end{equation}
respectively. The slab is assumed to be in
quasi-magnetohydrostatic equilibrium at all times, supported
against its own self-gravity by the magnetic and gas pressures.
The loss of flux from ambipolar diffusion is exactly compensated
for by the compression of the slab which is necessary to maintain
equilibrium. In this approximation, the left-hand side of equation
(\ref{momcon2}) is zero, and we may integrate the force balance
to obtain
\begin{equation}\label{momcon3}
  \frac{B^2}{2\mu_0} + a^2 \rho = 2\pi G (\sigma_\infty^2 - \sigma^2)
\end{equation}
where integration constant $\sigma_\infty$ is the value of
$\sigma$ at $z=\infty$ (where $\rho$ is zero).

Following the work of Shu~(1983), we introduce the non-dimension
quantities
\begin{eqnarray}\label{nond}
  \nonumber \tilde{\sigma} \equiv \frac{\sigma}{\sigma_\infty} , \quad
  \tilde{\rho} \equiv \frac{a^2}{2 \pi G \sigma_\infty^2} \rho,\quad
  \tilde{z} \equiv \frac{2 \pi G \sigma_\infty}{a^2} z,\quad
  \tilde{B} \equiv \frac{B}{2 \sigma_\infty \sqrt{\pi \mu_0 G}}~~ \\
  \tilde{t} \equiv (\frac{2\sqrt{2\pi G}}{\gamma_{AD} \epsilon}) (\frac{2\pi
             G\sigma_\infty}{a})t ,\quad
  \tilde{\epsilon} \equiv \frac{a^5}{(2\pi G)^{5/2}
  \sigma_\infty^5} \epsilon', \quad
  \tilde{v}_d \equiv \frac{\gamma_{AD} \epsilon}
  {a \sqrt{2 \pi G}} v_d,
\end{eqnarray}
so that we rewrite the basic equations (\ref{magcon2}),
(\ref{mascon2}) and (\ref{momcon3}) as follows:
\begin{equation}\label{magcon4}
  \frac{\partial}{\partial \tilde{t}}
  (\frac{\tilde{B}}{\tilde{\rho}})= \frac{\partial}{\partial
  \tilde{\sigma}}(\frac{\tilde{B}^2}{\tilde{\rho}^{1/2} + \tilde{\epsilon} \tilde{\rho}^{-2}} \frac{\partial
  \tilde{B}}{\partial\tilde{\sigma}}),
\end{equation}
\begin{equation}\label{momcon4}
  \tilde{B}^2 + \tilde{\rho}= 1- \tilde{\sigma}^2,
\end{equation}
\begin{equation}\label{mascon4}
  \frac{\partial \tilde{z}}{\partial
  \tilde{\sigma}}=\frac{1}{\tilde{\rho}},
\end{equation}
and the drift velocity (\ref{drift2}) as
\begin{equation}\label{drift3}
  \tilde{v}_d=-\frac{1}{\tilde{\rho}^{1/2} + \tilde{\epsilon} \tilde{\rho}^{-2}} \frac{\partial \tilde{B}^2}{\partial
  \tilde{\sigma}}.
\end{equation}
Here, a natural family of initial states is generated by assuming
that the initial ratio of magnetic to gas pressure is everywhere
a constant, $\alpha_0$, i.e., $\tilde{B}^2/\tilde{\rho} =
\alpha_0$ at $\tilde{t} =0$.  Then one finds from equations
(\ref{momcon4}), (\ref{mascon4}) and (\ref{drift3}) that
\begin{equation}\label{dens}
  \rho _{(z,t=0)} = \frac{\rho_0}{\cosh^2(z/z_\infty)},
\end{equation}
\begin{equation}\label{drift4}
  v_{d(z,t=0)} = \frac{2\alpha_0}{\sqrt{1+\alpha_0}} \frac{a \sqrt{2\pi G}}{\gamma_{AD} \epsilon}
  \frac{\sinh(z/z_\infty)}{1+\epsilon' \rho_0^{-5/2} \cosh ^5(z/z_\infty)},
\end{equation}
where $\rho_0 \equiv 2\pi G \sigma_\infty^2 / a^2 (1+\alpha_0)$ is
the central density of the slab at $t=0$, and $z_\infty \equiv a
\sqrt{(1+\alpha_0)/2\pi G \rho_0}$ is a length-scale parameter.

\begin{figure}
\centering
\includegraphics[width=3 in]{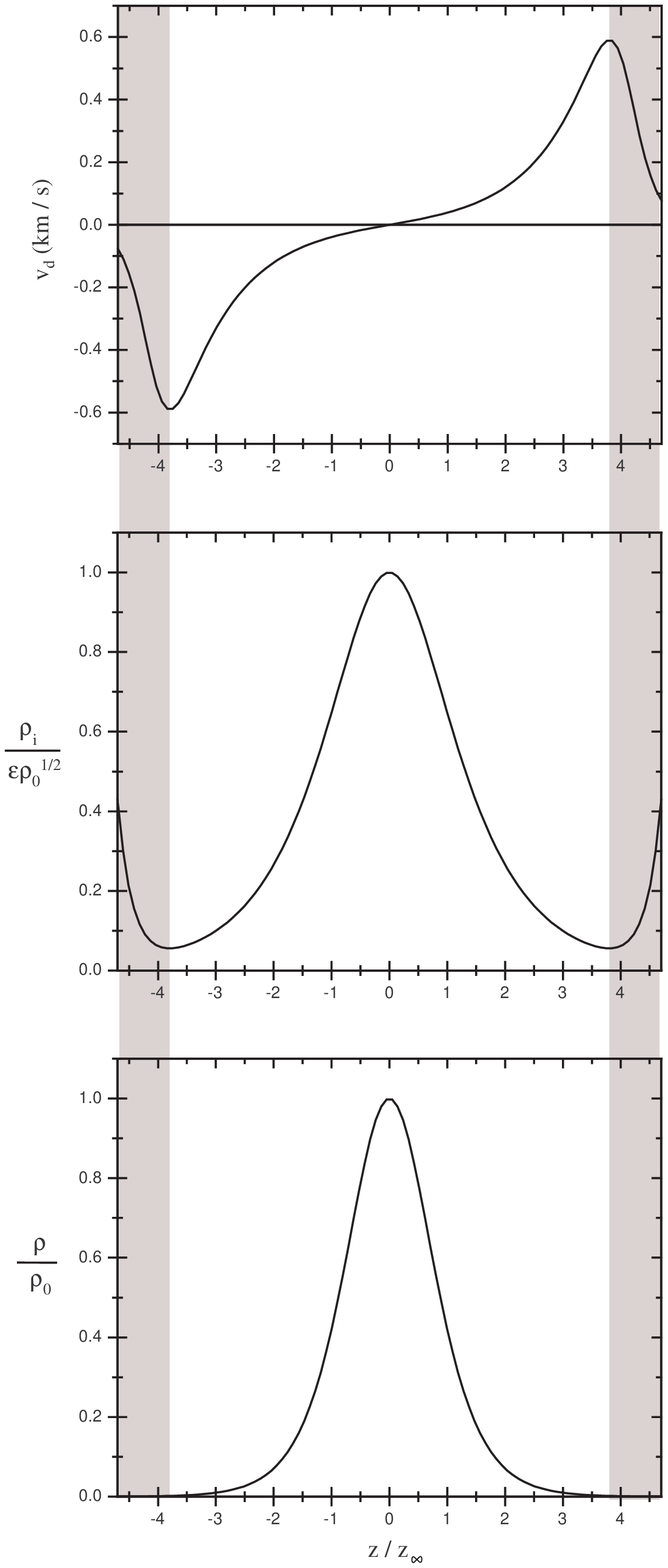}
\caption{The initial drift velocity, ion density, and neutral
density in the cloud and outercloud medium (gray region) for
$\rho_0 = 3.8 \times 10^{-15} \mathrm{kg.m^{-3}}$, $a=0.55
\mathrm{km.s^{-1}}$, and $\alpha_0=1$.}\label{cloincloud}
\end{figure}

Figure~\ref{cloincloud} shows the initial neural density, ion
density, and drift velocity in the cloud and outercloud medium for
$\rho_0 = 3.8 \times 10^{-15} \mathrm{kg.m^{-3}}$, $a=0.55
\mathrm{km.s^{-1}}$, and $\alpha_0=1$. Shu~(1983) solved the
equations (\ref{magcon4})-(\ref{drift3}) by finite difference
techniques under the initial and boundary conditions
\begin{equation}\label{bound}
  \tilde{B}= \left ( \frac{\alpha_0}{1+\alpha_0} \right )^{1/2}
  (1-\tilde{\sigma}^2)^{1/2} \;
  \mathrm{at} \; \tilde{t}=0 \quad \mathrm{and} \quad \tilde{B} =
  0 \; \mathrm{at} \; \tilde{\sigma}=1.
\end{equation}
The integrations are carried out to time $\tilde{t}=20$ with
initial conditions $\alpha_0=1$ and $10$, and the results are
presented graphically in his paper (Shu~1983). As the magnetic
field leaks from the neutral gas, the volume density of the
neutrals shifts in profile from equation (\ref{dens}) to that
case with $\alpha_0=0$, and the drift velocity gradually settles.

\section{Numerical scheme}

Since the molecular gas is expected to cool and heat rapidly,
considering the thermal energy is appropriate. For this aim, we
produce a two-fluid SPH implementation more efficient than HW, to
study the thermal evolution of the self-gravitating one
dimensional molecular cloud. In the technique of HW, the initial
SPH particles are represented by two sets of particles:
magnetized ion SPH particles and non-magnetized neutral SPH
particles. For each SPH particle we must create two separate
neighbor lists: one for neighbors of the same species and another
for those of different species. Consequently, each particle must
have two different smoothing lengths. In the following sections
we refer to neutral particles as $\alpha$ and $\beta$, and ion
particles as $a$ and $b$; the subscripts $1$ and $2$ refer to
both ions and neutral particles.

The neutral density in place of neutral particles is estimated via
usual summation over neighboring neutral particles
\begin{equation}
\rho_{n,\alpha}=\sum_\beta m_\beta W_{\alpha\beta},
\end{equation}
while in place of ions, $\rho_{n,a}$, is given by interpolation
technique from the values of nearest neighbors. The ion density
is evaluated via equation (\ref{ionden}) for both places of ions
and neutral particles. In the particle approximation of SPH
method, the infinitesimal volume in the integrations at the
location of particle $a$ is replaced by the finite volume of that
particle $\triangle V_a$, which is related to the mass and
density as $m_a = \rho_a \triangle V_a$. Since the ion density is
evaluated by equation (\ref{ionden}), no by the usual summation
rule
\begin{equation}\label{ionsum}
\rho_{i,a}=\sum_\beta m_b W_{ab},
\end{equation}
we must update the mass of ion $a$ as follows
\begin{equation}
m_a^{new}=m_a^{old} \frac{\rho_a^{new}}{\rho_a^{old}},
\end{equation}
in each time step so that the above summation/interpolation
routine for ions (\ref{ionsum}) being accurate.

The SPH form of the drift velocity of ion particle $a$ is given
by HW as
\begin{equation}\label{driftsph}
v_{d,a}= \frac{1}{\gamma_{AD}\rho_{n,a}}[-\frac{1}{\mu_0
\rho_{i,a}} \sum_b \frac{m_b}{\rho_{i,b}} B_b B_a
\frac{dW_{ab}}{dz_a} - \sum_b m_b \Pi_{ab} \frac{dW_{ab}}{dz_a}].
\end{equation}
where $\Pi_{ab}$ is the usual artificial viscosity between ion
particles $a$ and $b$ (Monaghan~1992). The artificial viscosity is
reconsidered in SPH to prevent inter-particle penetration,
unwanted heating, and unphysical solutions. Nejad-Asghar, Khesali
\& Soltani (2008) has recently considered the coefficients in the
Monaghan's standard artificial viscosity as time variable, and a
restriction on them is proposed such that avoiding the undesired
effects in the subsonic regions. Here, we use the Monaghan's
standard artificial viscosity, since the cloud contraction is
quasi-hydrostatic and there is not supersonic motions and shock
formation during this contraction. Keeping in mind the second
golden rule of SPH which is to rewrite formulae with the density
inside operators (Monaghan~1992), we can optimize the drift
velocity of HW as follows
\begin{eqnarray}\label{driftsph2}
\nonumber v_{d,a}= \frac{1}{\gamma_{AD}\rho_{n,a}}[-\frac{1}{2\mu_0
\rho_{i,a}} \sum_b \frac{m_b}{\rho_{i,b}} (B_b^2-B_a^2)
\frac{dW_{ab}}{dz_a}
\\ - \rho_{i,a} \sum_b \frac{m_b}{\rho_{i,b}} \Pi_{ab}
\frac{dW_{ab}}{dz_a}],
\end{eqnarray}
where two extra density terms are introduced, one outside and one
inside the summation sign. This comes as a result of the
approximation to the volume integral needed to perform function
interpolation. The drift velocity at neutral places are used to
estimate the drag acceleration
\begin{equation}\label{dragaccsph}
  a_{drag,\alpha} = \gamma_{AD} \rho_{i,\alpha} v_{d,\alpha},
\end{equation}
instead the method of HW who used the expression of Monaghan \&
Kocharayan~(1995). Since there is no any analytical expression
that allows us to calculate the value of drift velocity in place
of the neutral particles, we use the interpolation technique that
starts at the nearest neighbor, then add a sequence of decreasing
corrections, as information from other neighbors is incorporated
(e.g., Press et al. 1992).

In the usual symmetric form, the self-gravitating SPH
acceleration equation for neutral particle $\alpha$ is
\begin{equation}\label{accneut}
\frac{dv_\alpha}{dt}=g_\alpha-\sum_\beta m_\beta
(\frac{p_\alpha}{\rho^2_\alpha} +\frac{p_\beta}{\rho^2_\beta} +
\Pi_{\alpha\beta} )\frac{d W_{\alpha\beta}}{d z_\alpha}
+a_{drag,\alpha}
\end{equation}
where $g_\alpha$ is the gravitational acceleration of particle
$\alpha$. The ion momentum equation assuming instantaneous
velocity update so that we have
\begin{equation}
v_a= \sum_\beta \frac{m_\beta}{\rho_\beta} v_\beta W_{a\beta}
+v_{d,a}
\end{equation}
where the first term on the right-hand side gives the neutral
velocity field at the ion particle $a$, calculated using a
standard SPH approximation.

The SPH equivalent of the energy equation (\ref{energycon}) is
\begin{equation}
\frac{du_\alpha}{d t}=\frac{1}{2} \sum_\beta m_\beta
(\frac{p_\alpha}{\rho_\alpha^2} +\frac{p_\beta}{\rho_\beta^2}
+\Pi_{\alpha\beta}) v_{\alpha\beta} \frac{\partial
W_{\alpha\beta}}{\partial z_\alpha} -\Omega_\alpha.
\end{equation}
The temperatures of neutral particles are calculated from
equation~(\ref{ut}), and the temperature of any ion is assumed to
be same as its nearest neutral neighbor. Finally, the magnetic
induction equation (\ref{magcon}) in SPH form is replaced by
\begin{equation}
\frac{dB_a}{dt}= \sum_b \frac{m_b}{\rho_b} B_a v_{ab} \frac{d
W_{ab}}{dz_a},
\end{equation}
where the usual notations of the ion fluid are used.

\section{The computer experiments}

The chosen physical scales for length and time are $[l]=200
\mathrm{AU}$, and $[t]=10^3 \mathrm{yr}$, respectively, so that
velocity unit is approximately $[v]=1 \mathrm{km.s^{-1}}$. The
Newtonian constant of gravitation is set $G= 1 [m]^{-1} [l]^3
[t]^{-2}$ for which the calculated mass unit is $[m]= 4.5 \times
10^{29} \mathrm{kg}$. Consequently, the derived physical scale for
density, energy per unit mass, and drag coefficient are $[\rho]=
1.7\times 10^{-11} \mathrm{kg.m^{-3}}$, $[u]= 10^{6}
\mathrm{J.kg^{-1}}$, and $\gamma_{AD}=1.8\times 10^{10} [l]^3
[m]^{-1} [t]^{-1}$, respectively. In this manner, the numerical
values of $\epsilon$ and $\epsilon'$ are $1.8 \times 10^{-9}
[l]^{-3/2} [m]^{1/2}$ and $3.5 \times 10^{-17} [l]^{-15/2}
[m]^{5/2}$, respectively. The magnetic field is scaled in units such
that the constant $\mu_0$ is unity. Since the magnetic flux density
has dimensions
\begin{equation}\label{magdimen}
  [B]=\frac{[m]}{[t][charge]},
\end{equation}
while $\mu_0$ has dimensions
\begin{equation}\label{mudimen}
  [\mu_0]=\frac{[m][l]}{[charge]^2},
\end{equation}
specifying $\mu_0=1$ therefore scales the magnetic field equal to
$[B]=5.1 \mathrm{nT}$. With aforementioned units, the thermal energy
per unit mass (\ref{ut}) is represented by $8.3\times10^{-3}
(5X/4+3Y/8) T$, the heating rates due to cosmic rays and ambipolar
diffusion are $\Gamma_{CR} = 1.1 \times 10^{-3} [u]/[t]$ and
\begin{equation}\label{adheat}
  \Gamma_{AD,\alpha}= \gamma_{AD}\rho_{i,\alpha} v_{d,\alpha}^2,
\end{equation}
respectively, and the parameters for the gas cooling function are
\begin{eqnarray}\label{lambda02}
  \nonumber \log\left(\frac{\Lambda_\alpha}{[u]/[t]}\right) = -4.48 -
  0.87 (\log \frac{\rho_\alpha}{2.24\times10^{-4}}) \\ - 0.14 (\log \frac{\rho_\alpha}{2.24\times10^{-4}})^2,
\end{eqnarray}
\begin{eqnarray}\label{beta2}
 \nonumber \beta_\alpha = 3.07 - 0.11 (\log \frac{\rho_\alpha}{2.24\times10^{-4}})
  \\ - 0.13 (\log \frac{\rho_\alpha}{2.24\times10^{-4}})^2.
\end{eqnarray}

\subsection{Initial setting}

\begin{figure}
\centering
\includegraphics[width=4.5in]{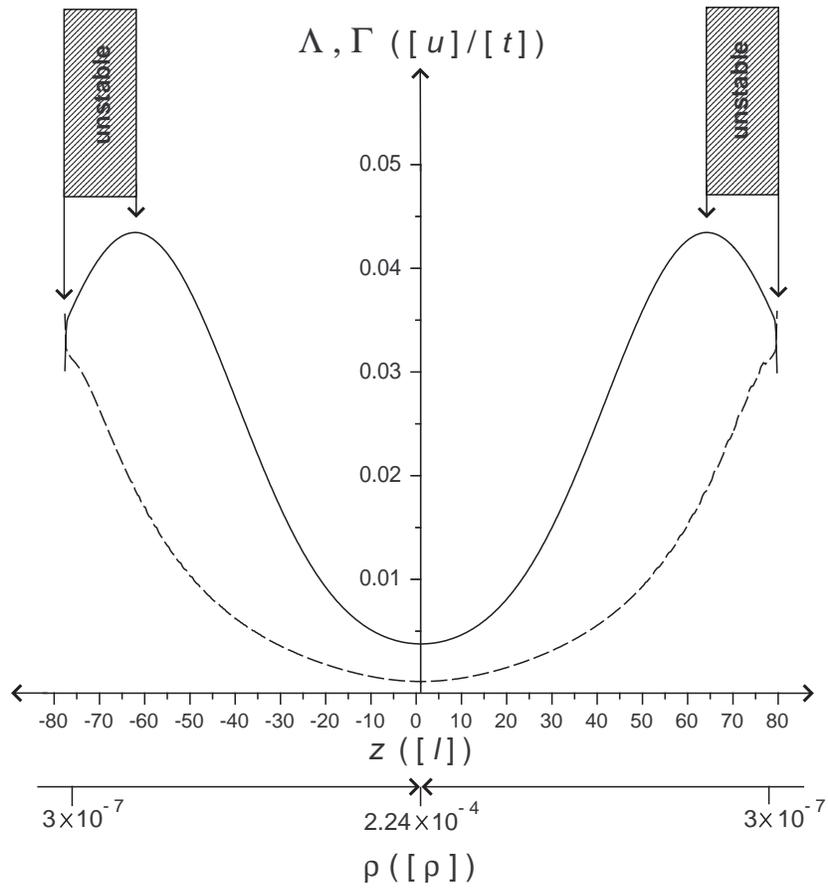}
\caption{The initial values of the cooling (solid) and heating
(dash) functions versus position and neutral density.}
\label{coolheat0}
\end{figure}

The initial conditions for this simulation are a parallel magnetic
field directed perpendicular to the $z$-axis so that the initial
ratio of magnetic to gas pressure is everywhere a constant
($\alpha_0=1$), and a density profile given by the equation
(\ref{dens}). The magnetic field is assumed to be frozen in the
fluid of charged particles and the central density is assumed to
be $\rho_0= 2.24\times10^{-4}[\rho]$. We choose a molecular cloud
which has a mass fraction of molecular hydrogen and helium
$X=0.75$ and $Y=0.25$, respectively, and has an initial uniform
temperature of $T_0 =50\mathrm{K}$. We assume that the cloud slab
is spread from $z=-78 [l]$ to $z=+78 [l]$ (according to
Fig.~\ref{cloincloud}). The initial values of the cooling and
heating functions are shown in Figure~\ref{coolheat0}. As
presented in this figure, the isobaric thermal instability
criterion,
\begin{equation}\label{thermcri}
  \frac{\partial \Lambda}{\partial \rho} > \frac{\partial \Gamma}{\partial
  \rho},
\end{equation}
is satisfied in the outer parts of the cloud, thus, these regions
are thermally unstable while the inner part is stable as outlined
by Nejad-Asghar~(2007).

Implementation of the boundary particles in the diffusion
processes is an important problem in two-fluid SPH simulation. In
ambipolar diffusion process, the ion particles are physically
diffused through the neutral fluid, thus, the ions will be bared
in the boundary regions of the cloud (i.e. without any neutral
particles in their neighbors). Both the cloud and boundary
regions contain ion and neutral particles. The complete system is
represented by $N$ discrete but smoothed SPH particles (i.e.
Lagrangian sample points) so that coagulation of particles and
fragmentation of the cloud is truthfully revealed. Since it is
desirable to have initially the same numerical resolution for
both fluid components, we use $N/2$ ions and $N/2$ neutral
particles. We set up boundary particles ($4h_1$ up and down in
$z$) using the linear extrapolation approach (from the values of
the inner particles) to attribute the appropriate drift velocity,
drag acceleration, pressure acceleration, energy rate, and the
magnetic induction rate to the boundary particles. We check the
position of ions before making a tree and nearest neighbor
search, so that we do not consider the bared ion particles in the
boundary regions of the simulation at next time-step.

The present SPH code has the main features of the TreeSPH class so
that the nearest neighbors searching are calculated by means of
this procedure. The selection of time-step, $\Delta t$, is of
great importance. There are several time-scales that can be
defined locally in the system. For each particle $1$, we
calculate the smallest of these time-scales using its smallest
smoothing length, $h_1$, i.e.
\begin{equation}\label{timestep}
\Delta t_1=C_{cour} \min[ \frac{h_1}{| v_1 |} , \frac{h_1}{v_{A,1}}
, \frac{h_1}{c_{s,1}}],
\end{equation}
where $v_A = B / \sqrt{\mu_0 \rho_i}$ is the Alfv\'{e}n speed of ion
fluid and $C_{cour}$ is the Courant number which in this paper is
adopted equal to $0.3$ (for numerical stability). The evolution were
carried out to time $<u> / <\Lambda,\Gamma>\sim 0.43/0.025= 17.2[t]$
so that the fragmentation of the cloud via thermal instability may
be revealed.

\subsection{Results}

\begin{figure}
\centering
\includegraphics[width=3.2in]{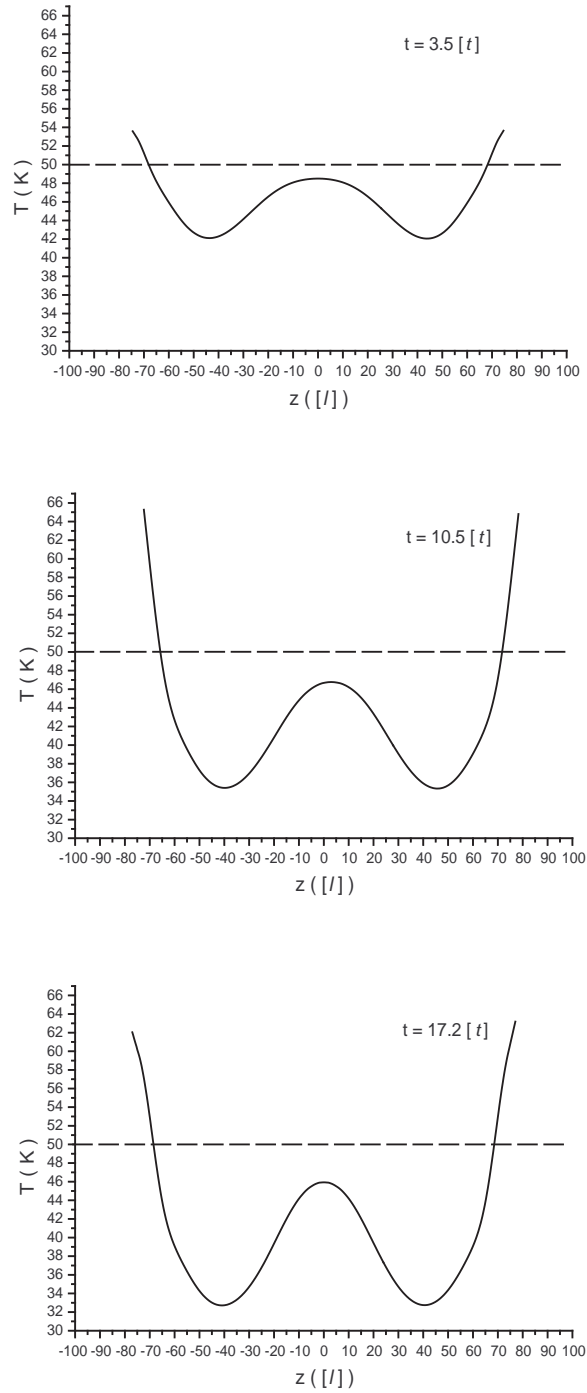}
\caption{The temperature profile versus position at times $t=3.5
[t]$, $10.5 [t]$ and $17.2 [t]$. The initial temperature of the
slab is chosen as uniform which is shown by dash line.}
\label{tempfig}
\end{figure}

The molecular cloud has an initial uniform temperature of $T_0=50
\mathrm{K}$. The Fig.~\ref{coolheat0} shows that the cooling is
greater than heating at initial time $t=0$, but by passing the
time, the drift speed of ions in the outer layers increases,
thus, the heating rate growths in those regions. Increasing of the
heating rate in the outer layers causes to stabilize those
regions, and the intermediate layers become thermally unstable.
The evolution of cooling and heating rates, causes to increase
the temperature of the slab at outer regions while decrease the
temperature at the intermediate parts of it. The temperature
profiles at times $t=3.5 [t]$, $10.5 [t]$ and $17.2 [t]$ are
shown in Figure~\ref{tempfig}. This figure shows that the outer
layers of the slab heat up, the intermediate layers cool down,
and the inner layers hardly change. It is obvious that the
instability of the cloud at the intermediate parts, causes the
formation of two relative cool regions in those areas.

\begin{figure*}
\centering
\includegraphics[width=6.5in]{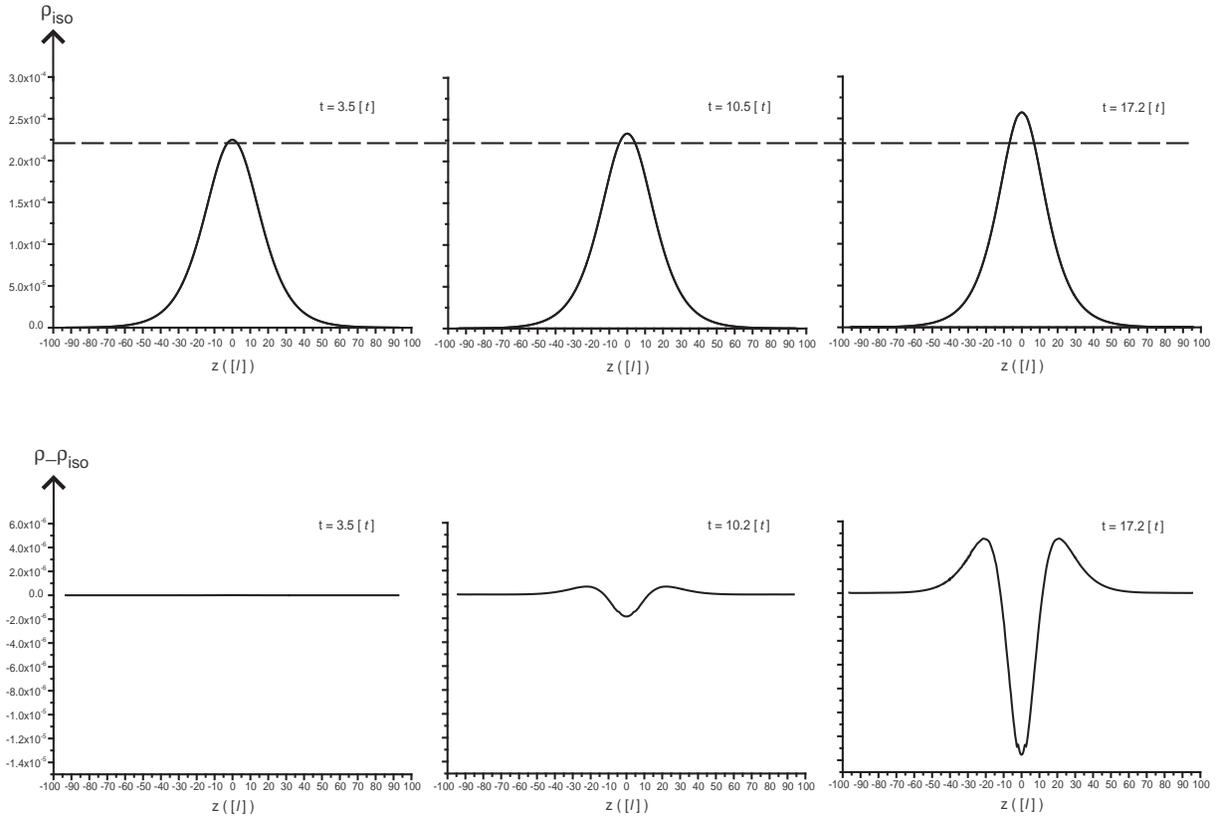}
\caption{The isothermal density profile versus position at times
$t=3.5 [t]$, $10.5 [t]$ and $17.2 [t]$ (top panel), and the
density comparison with the isothermal case at those times
(bottom panel). The dash line in the top panel is the initial
central density of the cloud.} \label{denfig}
\end{figure*}

Since, in course of time, the thermal instability is shifted to
the intermediate regions of the slab, the condensation mechanism
is occurred there as depicted in Fig.~\ref{denfig}. The growth of
thermal instability results a density imbalance between the
preformed condensations and their adjacent medium. As a result of
this instability process, the particles are dragged from the
adjoining of the condensations so that the density fluctuations
growth and the fragmentation of the parent slab into cold, dense,
low-mass cloudlets may be appeared. To compare the effect of the
thermal instability, we firstly perform the isothermal simulation
that have initially $T=50\mathrm{K}$. The thermal and density
evolution of the slab are carried out to times $t=3.5 [t]$, $10.5
[t]$ and $17.2 [t]$. The isothermal density profile and the
density comparison with the isothermal case, are shown in
Figure~\ref{denfig}. This figure shows the resulting isothermal
density evolution at the top panel while the density contrast,
relative to the isothermal case, is shown in the bottom panel. In
this run, the thermal instability is fully developed and makes
density fluctuations in the intermediate parts of the cloud.

\section{Summary and conclusions}

Molecular clouds have a hierarchical structure that extends from the
scale of the cloud down to much smaller masses for unbound
structures. A weakly ionized self-gravitating one dimensional slab
geometry is assumed in order to gain some insight into the
fragmentation process. The two-fluid SPH technique was used to
investigate the nonlinear thermal evolution of the slab. The initial
conditions for this simulation are an inverse square
cosine-hyperbolic profile for initial density and a parallel
magnetic field directed perpendicular to the slab.

The initial values of the cooling and heating functions are shown
in Figure~\ref{coolheat0}. According to this figure, the isobaric
thermal instability criterion is satisfied in the outer parts of
the cloud. In course of time, the thermal instability is shifted
to the intermediate regions of the slab so that the condensation
mechanism is occurred there, thus, these regions are thermally
unstable while the outer part is stable. The evolution were
carried out to time $17.2[t]$ so that the fragmentation of the
cloud via thermal instability is revealed.

The cooling and heating rates cause to increase the temperature of
the clouds at outer regions while decrease the temperature at the
inner part of it. It is obvious that the instability of the cloud
at the intermediate parts, causes the formation of two relative
cool regions in those areas, which is shown in
Figure~\ref{tempfig}. The rapid growth of thermal instability
results a density imbalance between the cloud and the
surroundings. The isothermal density profile and density
contrast, relative to the isothermal case, are shown in
Figure~\ref{denfig}. In this run, the thermal instability can
fully develop and makes density fluctuations in the intermediate
parts of the cloud.

This feature may be responsible for the planet formation in the
intermediate parts of a collapsing molecular cloud and/or may also
be liable for the formation of star forming dense cores in the
clumps. The macro-velocity fields in the molecular clouds are
highly turbulent and supersonic. It is then of uppermost
importance to consider both collision and merger of the formed
condensations. Merging is possibly the main onset mechanism to
form dense cores, which likely evolve to star formation.

\section*{Acknowledgments}

This work has been supported by Research Institute for Astronomy and
Astrophysics of Maragha (RIAAM).


\clearpage

\clearpage

\clearpage

\clearpage

\end{document}